\documentclass[graybox]{svmult}

\usepackage{amsmath}
\usepackage{mathptmx}       
\usepackage{helvet}         
\usepackage{courier}        
\usepackage{type1cm}        
\usepackage{makeidx}         
\usepackage{graphicx}        
\usepackage{multicol}        
\usepackage[bottom]{footmisc}

\newcommand{\ket}[1]{\mid #1 \rangle}
\newcommand{\w}{\omega}

\makeindex             

\begin{document}

\title*{Use Cases of Quantum Optimization for Finance}
\author{Samuel Mugel, Enrique Lizaso, and Rom\'{a}n Or\'{u}s}
\institute{Samuel Mugel \at Multiverse Computing,
Banting Institute, Toronto, ON M5G 1L5 Canada\\\email{sam.mugel@multiversecomputing.com}\and
Enrique Lizaso \at 
Multiverse Computing, Parque Cientifico y Tecnologico de Gipuzkoa, Paseo de Miram\'on 170, 20014 Donostia-San Sebasti\'an, Spain\\
\email{enrique.lizaso@multiversecomputing.com}\and
Rom\'an Or\'us \at 
Multiverse Computing, Parque Cientifico y Tecnologico de Gipuzkoa, Paseo de Miram\'on 170, 20014 Donostia-San Sebasti\'an, Spain\\
Donostia  International  Physics  Center,  Paseo  Manuel  de  Lardizabal  4, 20018  Donostia-San  Sebasti\'an,  Spain \\
Ikerbasque  Foundation  for  Science,  Maria  Diaz  de  Haro  3,  48013  Bilbao,  Spain \\
\email{roman.orus@multiversecomputing.com}}
%
\maketitle

\abstract{In this paper we briefly review two recent use-cases of quantum optimization algorithms applied to hard problems in finance and economy. Specifically, we discuss the prediction of financial crashes as well as dynamic portfolio optimization. We comment on the different types of quantum strategies to carry on these optimizations, such as those based on quantum annealers, universal gate-based quantum processors, and quantum-inspired Tensor Networks.}

\section{Introduction}

Some computational problems have no efficient classical solution. An important example is searching for an item in a disordered list. For a list of size $N$, this problem can be solved in $O(N)$ operations on a standard (\emph{classical}) computer. If you are searching Google's database for a given keyword, this can take a truly phenomenal amount of time. 

In some cases, quantum computing can solve intractable problems efficiently. A spectacular example is Grover's algorithm, which can find an item in a disordered list in $O(\sqrt{N})$ computations \cite{Grover1996}. There is no way to devise a classical algorithm with this performance.

The key to Grover's speedup is superposition. Quantum systems can be in several states at once. A quantum bit (\emph{qubit}), for instance, can be in a superposition of states $\ket{0}$ and $\ket{1}$. If we perform an operation on a system which is in many states at once, the operation is applied simultaneously to each state. Quantum computing therefore implements parallel computing natively on a massive scale.

At its core, a computation is a physical process. Quantum computing is the idea that quantum processes can also be used to perform calculations. The laws which govern quantum objects are different from the physical laws which are important for classical computing. This means that computational resources are available in quantum computing that don't exist in classical computing. There are a vast amount of examples where a computational advantage can be obtained from these resources \cite{nielsen_chuang_2010}.

There exist many problems of practical interest which cannot be solved efficiently on a classical computer. Many high value problems can be found in the field of finance. Option pricing, for instance, is usually solved through Monte Carlo due to the inherently stochastic nature of the variables involved. It is common for these computations to take upwards of 24 hours. Problems like credit scoring, financial threat detection, and customer identification usually employ complex machine learning models. Each of these can cost millions of dollars to train. There also exists a myriad of financial optimization problems of great monetary value. These include investment optimization, liquidity management, and macro-economics problems such as financial crash prediction. These problems are complex because financial institutions or assets can be correlated in extremely complex ways. Optimization problems are often the bottleneck of financial calculations.

There is an intense research effort underway to efficiently solve high value financial problems using quantum computing. This field has already demonstrated spectacular applications of quantum computing of commercial value, several of which are reviewed in this paper. For a complete description of verticals of quantum computing in finance, see Ref. \cite{Orus2018}.

In this article, we focus on intractable financial optimization problems, and review some examples of how these can be solved efficiently using quantum computing.  In Sec. \ref{sec:financial-crash}, we describe an algorithm to find the stable state of a financial network using a present day quantum computer. This result can be used for financial crash prediction. In Sec. \ref{sec:portfolio-optimization}, we find the optimal portfolio of investment, chosen from 52 assets over a four year period. This work relied on present day quantum computers and original quantum inspired methods.

\section{Predicting Financial Crashes}
\label{sec:financial-crash}

We live in a world where financial institutions hold vast portfolios, which often include shares in other financial institutions. In this situation, \emph{could a particular transaction lead to a financial crash}?

This is an incredibly complex question. It's so complex that traditional supercomputers are incapable of tackling it, even for simple toy models. Even given complete knowledge of all the assets and cross-holdings in a simple network of 20 institutions, it would take more time than the age of the universe (13.7 billion years!) to compute the effect of a perturbation.

At present, we mainly rely on empirical or statistical tools to answer this question. It is not clear that these methods can systematically and reliably predict financial crashes because indicators of a crisis generally fail at predicting the next crisis. Our failure to prevent these events is directly responsible for economic crises and their devastating consequences.

While the situation might seem dire, quantum computing has shown it could tackle these types of problems efficiently, both in theory and in practice. In particular, quantum computing has proved extremely successful at tackling complex financial problems \cite{Orus2018}.

The work presented in this section builds upon Ref. \cite{Orus2018a}. A physical implementation of this work was presented in \cite{Ding2019}.

\subsection{Quantum Annealing}
\label{sec:quantum_annealing}

In Ref. \cite{Orus2018a}, we developed a quantum algorithm for predicting financial crashes on a quantum annealer. These quantum processors solve quadratic integer valued optimization problems, which can be intractable for classical computers.

To solve a problem on a quantum annealer, we map the problem's cost function to an idealized quantum system, the Ising model. The Ising model analogous to our problem is simulated on the quantum annealer. The annealing procedure drives the simulated material to its most stable state, which can then be read off \cite{Kadowaki1998}.

Here, the cost function is the stability equation of the financial network. Its most stable state -- or \emph{ground state} -- is the network's equilibrium configuration. When the network is perturbed, we can measure the new equilibrium of the system and ask: has a financial crash occurred to reach this configuration?

\subsection{Financial Model}

We base our study on a model for financial networks originally proposed by Elliott, Golub, and Jackson \cite{Elliott2014}. Their model predicts the equilibrium \emph{market value} of an institution, which is defined as its non-inflated value to the economy. The vector of equilibrium market values $\vec{v}$ satisfies the financial stability condition:
\begin{equation}
\label{eq:stability_equation}
 \vec{v} = A \left(D\vec{p} - \vec{b}(\vec{v}, \vec{p})\right).
\end{equation}
From this equation, we can see that the market value of an institution is determined by two terms. First, it is determined by the price $p_j$ of each asset $j$, scaled with the asset ownership $D$. This matrix has terms $D_{ij}$, the share of asset $j$ held by the i$^{th}$ institution. The asset prices are related to the institutions' market values through the dependency matrix $A$, which accounts for institutions' cross ownership and debt contracts between institutions.

The second term $\vec{b}(\vec{v}, \vec{p})$ is a discontinuous quantity. It has value $b_i = 0$ when $v_i$ is above a particular value $v_i^c$. For $v_i^c<v_i^c$, the i$^{th}$ institution suffers an extra drop in value $b_i$. This term captures sociological effects, such as investors' loss of faith, and the loss of capital incurred when a company goes into liquidation.

It is straightforward to map Eq. (\ref{eq:stability_equation}) to an optimization problem:
\begin{equation}
\label{eq:crash_cost}
\vec{v} = \text{argmin}\vec{_v} \left\{ \vec{v} - A \left[D\vec{p} - \vec{b}(\vec{v}, \vec{p})\right] \right\}^2.
\end{equation}
Because Eq. \eqref{eq:crash_cost} is a discontinuous equation, it is extremely difficult to minimize.

\subsection{Formulation as a QUBO}
\label{sec:qubo_formulation}

To solve Eq. \eqref{eq:crash_cost} on a quantum annealer, we encode the market values $\vec{v}$ on our quantum processor. One way to do this is to use a binary encoding, such that institution $i$'s market value is given by:
\begin{equation}
\label{eq:binary_encoding}
v_i = \sum_q 2^q x_{iq},
\end{equation}
where $x_{iq}\in\{0,1\}$ is the readout value of the $q^{th}$ qubit assigned to institution $i$.

All the details of the financial network are encoded as interactions between qubits. These are captured in the Hamiltonian $H$:
\begin{equation}
\label{eq:Hamiltonian}
H = \left\{ \hat{\vec{v}} - A \left[D\vec{p} - \vec{b}(\hat{\vec{v}}, \vec{p})\right] \right\}^2.
\end{equation}
Note that we have defined the operator $\hat{\vec{v}}$ by quantizing Eq. \eqref{eq:crash_cost}. This means we have replaced the value variables $\vec{v}$ by operators -- mathematical objects which act on the quantum processor's qubits and determine their state (0, 1, or some superposition of these states).

In Ref. \cite{Ding2019}, we implemented Eq. \eqref{eq:Hamiltonian} on a physical quantum annealer. Because all present day quantum annealers can only solve quadratic optimization problems -- \emph{QUBOs} -- we had to approximate Eq. \eqref{eq:Hamiltonian}. We used ancilla qubits to simulate the complex, multi-qubit interactions between the logical qubits. In general, we can approximate the lowest energy state of an interaction between $k$ qubits by using $k$ ancillary qubits \cite{Chancellor2017}.

The next step is to enter the approximate $H$ into the quantum annealer, which finds its most stable state. It is this state that encodes the values at equilibrium of all assets in the financial network. The rest is easy: we calculate the values of all institutions, and ensure that none of them drops below their critical value. In the future, this could allow us to proceed responsibly, ensuring that none of our actions can cause financial havoc.

\section{Portfolio Optimization}
\label{sec:portfolio-optimization}

Investors are always on the lookout for the best investment opportunities. In general, this is a complex sequence of buys and sells. What makes this problem hard is that every buy or sell of shares has cost. This means that the optimal investment at time $t$ is highly dependent on your past investment trajectory.

Excellent tools exist, which exploit the cost function structure to tackle large optimization problems. This is for instance the case of gradient descent methods. Unfortunately, financial optimization problems are in general non-convex, meaning they do not have a simple mathematical structure. This makes them very hard to solve using traditional computing methods.

What makes our problem even harder is this: we only allow our investor to invest in large bundles. This is typically the case for exchange-traded funds (ETF) shares. Time can also be discrete. You may, for instance, choose to buy and sell assets only once a day. This helps avoid solutions with many short transactions, as these generally lead to highly fluctuating (i.e: risky) portfolios.

This means the variables we are optimizing -- the amount of assets you own -- are integer variables. The fact these are not continuous means we cannot use standard calculus tools, making it almost impossible to tackle this problem using classical computers. Clearly this is a high value problem where new, adapted optimizers could prove disruptive.

In Ref. \cite{Mugel2020}, we described a way to tackle this problem using quantum and quantum inspired methods. These tools allowed us to find the optimal investment trajectory spanning 52 assets and four years of data. This study is important because it is one of the few applications of quantum computing to finance of commercial value.

Financial optimization is an active field of study, which has given rise to many exciting studies, which include Refs. \cite{Rosenberg2016,Elsokkary2017a,Grant2020,Cohen2020}.

\subsection{Optimizers' Overview}

The solver treats the objective function like a black box. The solver gives the box a portfolio, the box returns its score. The aim for the solver is to repeat this process until the best portfolio is found. A good solver repeats the process as few times as possible.

Financial optimization problems are difficult because there is an almost infinite number of portfolios you could build. However not all of these are interesting to us: most portfolios have mediocre performance.

\subsubsection{Quantum Annealing}

One optimization method we implemented is quantum annealing, which we already described in Sec. \ref{sec:quantum_annealing}. We used it to explore only portfolios which are close to optimal, while ignoring the ones which are of no interest to us. This is analogous to the way water spreads over land, exploring all the valleys but none of the peaks.

\subsubsection{Variational Quantum Eigensolver (VQE)}

While solving an optimization problem, your solver must jump from one solution to another. Quantum computers have an advantage here because, in the quantum world, there exist many more ways to go from one solution to the next. While these intermediate portfolios have no physical significance, they can give the cost function a simpler (convex) structure. This allows us to use popular optimization tools such as gradient descent. This is the idea behind the Variational Quantum Eigensolver (VQE) algorithm. 

The quantum processor (QPU) is necessary in the VQE algorithm because an exponential number of parameters are needed to describe the intermediate portfolios. This makes it very expensive for a classical computer to calculate the value of the cost function. We therefore differ this task to the quantum processor. The classical computer is used to iteratively prepare portfolios, and find the lowest cost solution. 

\subsubsection{Tensor Networks}

In some cases, a smart use of classical computers can perform just as well as a quantum computer. By extending the space of possible portfolios, we can find shortcuts to the optimal solution.

However, not all of these new intermediates are important to find the shortcut. Tensor networks are a family of algorithms which simulate quantum mechanics on a classical computer, focusing only on the subspace of relevant portfolios. 

Using these algorithms, we were able to build a solver which can compete with -- and in some cases outperform! -- our quantum solver.

\subsection{The Financial Model}

According to Modern Portfolio Theory, a portfolio's risk can be estimated from its volatility, which is related to its covariance. The optimal investment can be found by maximizing the portfolio's profit, penalized with risk \cite{Singleton2018}. The ratio of risk to returns is the \emph{Sharpe ratio}, which is our metric for comparing investments.

The optimal amount $\w_{tn}$ of asset $n$ held at time $t$ can then be found by minimizing:
\begin{equation}
\label{eq:bare_cost_portfolio}
H_0 =  \sum_{t} 
 -\mu_t^T \w_t
 + \frac{\gamma}{2} \w_t^T \Sigma_t \w_t
+ \lambda (\Delta \w_t)^2. 
\end{equation}
Here, $\mu_t$ are the returns at time $t$, $\Sigma_t$ is the covariance matrix, and $\lambda$ is the matrix of transaction costs. The \emph{risk aversion} $\gamma$ is the portfolios penalty for risk. This determines the amount of risk an investor is willing to take.

The $\w_t$ vector should be normalized, meaning that $100\%$ of the available budget is invested at any time. We enforced this by penalizing portfolios which did not respect this constraint. The cost function we optimized was therefore:
\begin{equation}
\label{eq:cost_portfolio}
H = H_0 + \sum_t \rho \left(\sum_n \w_{tn} - 1 \right)^2,
\end{equation}
where $\rho$ is a hyperparameter of the model.

Note that Eq. \eqref{eq:cost_portfolio} is already in quadratic form. Using the procedure described in Sec. \ref{sec:qubo_formulation}, it is straightforward to map this to an Ising Hamiltonian.

\subsection{Dimensional Reduction Algorithms}

The largest quantum computer on Earth only has 5000 noisy quantum bits (for reference, a standard laptop has billions of bits). How can we fit such a large problem on such a small chip?

\subsubsection{Asset Clustering}

One way to reduce the problem's size is to group assets which move in similar ways. The key intuition here is that investing in highly correlated assets does not significantly lower the portfolios risk. A rough description of the optimal portfolio can be gained by grouping similar assets as indices. Once an initial guess is obtained, it can be used as a starting point for a finer portfolio search.

This is not only useful when resources are scarce. Even the best solvers struggle when the space of candidate portfolios is vast. Eliminating uninteresting portfolio options early reduces the risk of getting stuck with a suboptimal solution.

\subsubsection{Problem Fragmentation}

In general, you can't just invest in the most profitable shares every day: the costs of buying and selling stocks will likely outweigh your profits. You can expect, however, the best investment portfolio for one month to be some combination of each day's near optimal investments. Because quantum processors are samplers, it is easy for them to give us many good portfolios. These can then be combined using classical computers (which is computationally cheap).

This is an example of a hybrid algorithm. We fragmented our problem into two bits: 1) find several good investments for today; 2) combine them into a portfolio. We gave the quantum computer small, hard jobs. This means its small number of quantum bits wasn't an issue. We brought all these results together using a classical computer. This makes the most of its huge computational resources, without being bogged down by its slow algorithms.

\begin{table}
\centering
\begin{tabular}{l | c  c  c  c  c  c} 
& ~~~XS~~   & S   & M   & L   & XL   & XXL   \\ 
\hline
   $N_a$   &   3   & 4 &   4   &   8   & 8 &   8   \\ 
   $N_t$   &   2   & 5 &   7   &   17   & 29 &   53   \\ 
   $N_q$  &   1   & 1 &   1   &   2   & 2 &   3   \\ 
   $N_{\text{qubits}}$   &   6   & 20 &   28   &   272   & 464 &   1272   \\ 
   $N_{\text{states}}$    &   $64$   & $O(10^6)$ &   $O(10^8)$   &   $O(10^{81})$   & $O(10^{139})$ &   $O(10^{382})$   \\ 
   $K$   &   2   & 3 &   3   &   5   & 10 &   15   \\ 
   $K'$ &   1   & 1 &   1   &   3   & 3 &  7  \\ 
\end{tabular}
\caption{Specifics of the different datasets used for benchmarking the different algorithms and hardware platforms. Risk aversion for all datasets is fixed to $\gamma = 1$. Time steps $N_t$ are measured in business months (i.e., not considering weekends due to closure of some markets). $N_a$ is the number of assets, $N_q$ the number of bits per asset, $N_{\text{qubits}}$ the total number of qubits, $N_{\text{states}}$ the total number of sates in the portfolio space, $K$ the total amount of investment, and $K'$ the maximum investment per asset.}
\label{tab:sets}
\end{table}
\begin{table}
\centering
\setlength{\tabcolsep}{0.5em}
\begin{tabular}{l | c c c c c c} 
Method & XS & S & M & L & XL & XXL \\
\hline
VQE&3.59&-&-&-&-&-\\ 
Exhaustive&6.31&8.90&-&-&-&-\\ 
 VQE  Constrained &6.31&6.04&4.81&-&-&-\\ 
Gekko&5.98&8.90&8.39&15.83&20.76&-\\ 
 D-Wave  Hybrid &5.98&8.90&8.39&7.47&9.70&12.16\\ 
 Tensor  Networks &5.98&8.90&9.54&16.36& 15.77 & 15.83\\ 
\end{tabular}
\caption{Sharpe ratios obtained using the different optimization algorithms considered in Ref. \cite{Mugel2020}. The dataset parameters are described in Table \ref{tab:sets}.}
\label{tab:sharpe}
\end{table}

\subsection{Results}

Using quantum computers and tensor networks, we were able to find the best portfolio among $10^{382}$ candidates (see Table \ref{tab:sets}) -- many, many times more than the number of atoms in the Universe! This problem is completely out of reach of standard computers. Our final portfolio boasted a staggering $68\%$ return on investment over a 4 year period. A complete benchmarking of the solutions, as compared by the Sharpe ratio, is shown Table \ref{tab:sharpe}.

\section{Conclusions}

In this paper we have sketched two basic applications of quantum optimization for financial problems, namely, financial crash prediction, and dynamic portfolio optimization. These examples show that real business value can be derived from present day quantum computers. This is particularly true for the portfolio optimization case, where we found the best investment portfolio by optimizing over 52 assets and four years of data.

There exist many more verticals of quantum optimization in finance. It is our belief that the most interesting applications have yet to be discovered. It is time for financial institutions to revisit bottleneck problems. Quantum computing has already proved it can break down computational barriers, and is redefining what we consider \emph{intractable}.

\section*{Acknowledgements}

Thanks to Christophe Jurczak, Pedro Luis Uriarte, Pedro Mu\~{n}oz-Baroja, Creative Destruction Lab, BIC-Gipuzkoa,  DIPC, Ikerbasque, and Basque Government for constant support. We extend special thanks to our collaborators Francesco Benfenati, Be\~{n}at Mencia Uranga, and Samuel Palmer, for stimulating discussions and interesting ideas.

\bibliography{bibliography}

\end{document}